\documentclass[a4paper,11pt]{article}
\pdfoutput=1

\usepackage{amssymb,amsmath,bm}
\usepackage{a4wide}
\usepackage{color}
\usepackage{slashed}
\usepackage{graphicx}
\usepackage{amsfonts}
\usepackage{lscape}
\usepackage{hyperref}
\usepackage{amsthm}
\hypersetup{
    colorlinks=true, 
    linktoc=all,     
    linkcolor=blue,  
    citecolor= black
}
\usepackage{booktabs}
\usepackage{array}
\usepackage{rotating}
\usepackage[numbers, sort&compress]{natbib}
\usepackage{float}
\usepackage[utf8]{inputenc}
\usepackage[T1]{fontenc}
\newcommand{\GeV}{{\, \rm GeV}}

\newcommand{\eps}{\epsilon}

\newcommand{\be}{\begin{equation}}
\newcommand{\ee}{\end{equation}}

\newcommand{\cref}[1]{Chapter~\ref{ch:.#1}}

\newcommand{\beq}{\begin{equation}} 
\newcommand{\eeq}{\end{equation}} 
\newcommand{\ba}{\begin{array}}  
\newcommand{\ea}{\end{array}} 
\newcommand{\bea}{\begin{eqnarray}}  
\newcommand{\eea}{\end{eqnarray} }  
\newcommand{\bal}{\begin{align}}
\newcommand{\eal}{\end{align}}   
\newcommand{\bi}{\begin{itemize}}  
\newcommand{\ei}{\end{itemize}}  
\newcommand{\ben}{\begin{enumerate}}  
\newcommand{\een}{\end{enumerate}}  
\newcommand{\bc}{\begin{center}}
\newcommand{\ec}{\end{center}} 
\newcommand{\bt}{\begin{table}}
\newcommand{\et}{\end{table}}  
\newcommand{\btb}{\begin{tabular}}
\newcommand{\etb}{\end{tabular}}

\definecolor{mypink}{RGB}{219, 48, 122}

\begin{document}

\vspace{1cm}
\begin{titlepage}
\vspace*{-1.0truecm}
\begin{flushright}
CERN-TH-2018-036 \\
TTP18-012 \\
 \vspace*{2mm}
 \end{flushright}
\vspace{0.8truecm}

\begin{center}
\boldmath

{\Large\textbf{
A Grand-Unified Nelson-Barr Model
}}
\unboldmath
\end{center}

\vspace{0.4truecm}

\begin{center}
{\bf Jakob  Schwichtenberg$^a$, Paul Tremper$^a$, Robert Ziegler$^{a,b}$}
\vspace{0.4truecm}

{\footnotesize

$^a${\sl Institute for Theoretical Particle Physics, 
Karlsruhe Institute of Technology, \\ Engesserstrasse 7, D-76128 Karlsruhe, Germany \vspace{0.2truecm}}

$^b${\sl Theoretical Physics Department, CERN, 1211 Geneva 23, Switzerland \vspace{0.2truecm}}

}
\end{center}

\begin{abstract}
\noindent 
We argue that the Nelson-Barr solution to the Strong CP Problem can  be naturally realized in an E$_6$ Grand-Unified Theory. The chiral SM fermions reside in three generations of E$_6$ fundamentals together with heavy vectorlike down quarks, leptons doublets and right-handed neutrinos. CP is imposed on the Lagrangian and broken only spontaneously at high scales, leading to a mixing between chiral and vectorlike fields that allows to solve the Strong CP Problem through the Nelson-Barr mechanism. The main benefit of the E$_6$ GUT structure is the predictivity in the SM fermion sector, and a perfect fit to all SM observables can be obtained despite being over-constrained. Definite predictions are made for the neutrino sector, with a Dirac CP phase that is correlated to the CKM phase, allowing to test this model in the near future. 
\end{abstract}

\end{titlepage}

\newpage

\renewcommand{\theequation}{\arabic{section}.\arabic{equation}}


\section{Introduction}
\setcounter{equation}{0}

One of the most puzzling aspects of the Standard Model (SM) is the absence of CP violation in strong interactions, as measured by the topological angle 
\be
\bar\theta=\theta_{\rm QCD}-\theta_{\rm F} \, ,
\ee
where $\theta_{\rm QCD}$ denotes the coefficient of $\alpha_s^2/ 8 \pi \, G\tilde G$ and $\theta_{\rm F} = \arg \det M_u M_d$.  From the contribution to the neutron electric dipole moment one finds 
$\overline{\theta} <10^{-10}$ from the 95\% CL bound $|d_n| \le 3.6 \times 10^{-26} e \, {\rm cm}$~\cite{nEDM}, although generically one would expect $\overline{\theta}$ to be of the order of the observed CP violating phase in weak interactions, i.e. $\overline{\theta} \sim {\cal O}(1)$. Indeed it is the presence of the large CKM phase that prevents to forbid $\overline{\theta}$ by imposing CP as a fundamental symmetry.

The most popular explanation for this puzzle is the Peccei-Quinn mechanism~\cite{Peccei:1977ur,Peccei:1977hh}, which has the axion as a low-energy remnant~\cite{WW1, WW2}. This prediction makes axion models testable in upcoming experiments, which search for the axion with haloscopes like ADMX~\cite{ADMXfuture}, helioscopes like IAXO~\cite{IAXO1,IAXO2} or even precision flavor experiments like NA62~\cite{NA621,NA622, Japs, Axiflavon, Astrophobic}.

An alternative explanation for the smallness of $\overline{\theta}$ is provided by the Nelson-Barr mechanism \cite{Nelson1,Barr1,Nelson2,Barr2}, where CP is broken spontaneously at high scales. The original Lagrangian is CP invariant and hence $\theta_{\rm QCD}$ is zero. CP is broken spontaneously by large vacuum expectation values (VEVs), and CP violation is mediated to the low-energy Lagrangian only via mixing with heavy vectorlike quarks. If the Lagrangians respects two simple conditions (the so-called Barr-criteria), the resulting SM quark mass matrices are complex but have a real determinant, thus providing the CKM phase but rendering $\theta_F =0$ at tree-level. Finite and calculable contributions to $\theta_F$ arise at loop-level, but are generically suppressed by small Yukawa couplings and/or small mass ratios~\cite{Nelson1, Nelson2, Barr2, Vecchi}. The general Nelson-Barr framework has been realized in a minimal setup in Refs.~\cite{Branco1, Branco2}, and recently been combined with the idea of cosmological relaxation~\cite{Relaxion} in Ref.~\cite{Diego}. 

In contrast to axion models, in Nelson-Barr scenarios the effective theory below the scale $V_{\rm CP}$ of spontaneous CP breaking is just the SM. This scale is in general required to be very large  in order to  suppress loop corrections to $\overline{\theta}$ that are proportional to $v^2/V_{\rm CP}^2$~\cite{Branco1, Branco2}. Since $V_{\rm CP}$ sets the scale of the heavy vectorlike fermions, they are too heavy to be observed in the near future. Therefore the main drawback of Nelson-Barr models is the {\it lack of predictivity}, in addition to theoretical shortcomings discussed in e.g. Ref.~\cite{Dine}. 

In this paper, we address the issue of predictivity by embedding the Nelson-Barr mechanism into an E$_6$ Grand-Unified framework. This allows to connect the phases in the neutrino sector to the CKM phase, and  in particular to predict the Dirac CP phase that will be measured in the near future. Indeed the heavy vectorlike quarks needed in the Nelson-Barr setup naturally find their theoretical motivation in Grand-Unified theories (GUTs), as proposed already in Ref.~\cite{Barr2}. Among the possible simple GUT groups, $E_6$ \cite{Gursey:1975ki,Shafi:1978gg,Stech:1980fn,Barbieri1980369,  Borut3}  is ideally suited for the implementation of the Nelson-Barr mechanism, because the fundamental representation of $E_6$ contains in addition to chiral SM fermions a vectorlike pair of right-handed (RH) down quarks, besides a vectorlike pair of left-handed (LH) leptons and two RH neutrinos. Spontaneous CP breaking will induce a mixing between these vectorlike fields with the chiral fermions, and complex phases will enter low-energy quark, charged lepton and neutrino masses in a correlated manner. Definite predictions in the neutrino sector are then possible because of the very restricted form of the fundamental Yukawa sector, imposed by the E$_6$ GUT structure together with spontaneous CP violation. While in usual GUT scenarios the unification of Yukawa couplings is often problematic for light fermion generations, it turns out that the mixing with the heavy vector-like fields allows to cure these problems and to obtain a perfect fit to the full SM fermion sector. 

Therefore in our model the Nelson-Barr mechanism becomes predictive in the neutrino sector because of the E$_6$ GUT structure, which in turn is  phenomenologically viable because of the mixing with the heavy fermions needed to generate the CKM phase.

The rest of this paper is organized as follows: in Section 2 we present the general setup of the model and derive analytical expressions for the low-energy quark, charged lepton and neutrino masses. In Section 3 we perform a numerical fit to fermion masses and mixings and demonstrate that a perfect fit can be obtained for all observables with definite predictions for the neutrino sector. In Section 4 we discuss loop corrections to $\overline{\theta}$, which will constrain the overall scale of spontaneous CP breaking that is left undetermined by the fit. We finally summarize and conclude in Section 5. 

\section{An $E_6$ Nelson-Barr Model}

We embed the SM fermions in three E$_6$ fundamentals ${\bf 27}_i$ that decompose under SU(5) as 
\begin{align}
{\bf 27} = ({\bf 10} + {\bf \overline{5}} + {\bf 1})_{\bf 16} + ({\bf 5} + {\bf \overline{5} })_{\bf 10} + {\bf 1}_{\bf 1} \, ,
\label{eq:27}
\end{align}
where the subscripts denote the SO(10) decomposition. Thus for each generation of chiral SM fermions residing in the ${\bf 10_{16}}$ and ${\bf \overline{5}_{16}}$, there is a vectorlike pair of RH down-quarks and LH lepton doublets contained in  $({\bf 5 + \overline{5})_{10}}$ and two SM singlets ${\bf 1_{16}}$ and ${\bf 1_1}$. 

The vectorlike $({\bf 5 + \overline{5})_{10}}$  pair will get a large mass  at an intermediate scale $M \sim 10^9 \GeV$, and a mass term of similar order that mixes the heavy fermions in the ${\bf \overline{5}_{10}}$ with the chiral RH down quark and LH charged leptons in the ${\bf \overline{5}_{16}}$. According to the Nelson-Barr mechanism, this mixing is the only way how a complex phase enters  the low-energy effective (down) Yukawa couplings, which are of the form $y_d \sim y \cdot a$, where $y$ is a real and $a$ a hermitian $3 \times 3$ matrix. Indeed this matrix has a physical phase while the determinant stays real. 

The SM singlet ${\bf 1_1}$ will acquire a mass at the GUT scale $M_{\rm GUT} \sim  10^{16} \GeV$ from $E_6$ breaking, while the other singlet ${\bf 1_{16}}$ gets a mass at an intermediate scale $M_\nu \sim  10^{11} \GeV$ and induces neutrino masses via the Type-I seesaw mechanism. 

All fermion mass terms and the breaking of $E_6$ to the SM gauge group arise from adding scalars in ${\bf 27}_H$, ${\bf 351^\prime}_H$ and ${\bf 78}_H$ that develop large VEVs. The latter field is only responsible for breaking SO(10) at $M_{\rm GUT}$, while the other two fields couple to fermions according to the $E_6$-invariant Yukawa Lagrangian
\begin{align}
{\cal L}_{\rm yuk} & = {\bf 27}_ i {\bf 27}_j \left( Y_{27, ij} {\bf 27}_H + Y_{351^\prime,  ij} {\bf 351^\prime}_H \right) +{\rm h.c.} 
\label{L1}
\end{align}
We impose CP as a symmetry of the Lagrangian, so that the couplings $Y_{27}$ and $Y_{351^\prime}$ can be taken as real and symmetric $3 \times 3$ matrices. Without loss of generality we can choose the flavor basis such that $Y_{351^\prime}$ is diagonal. The E$_6$ structure therefore drastically reduces the number of flavor sector parameters, so that there are just 3+6 real parameters responsible for generating masses and mixings for quarks, charged leptons and neutrinos. While in usual GUT models such a unified structure often prevents to correctly account for all mass hierarchies, it turns out that in our setup the additional mixing in the RH down and LH charged lepton sector allows for an excellent fit to the full set of SM masses and mixings, as we are going to see in the next section.   

We do not spell out the scalar potential, which is simply assumed to generate the appropriate VEVs  and make all physical scalars except the SM Higgs ultra-heavy, around the scale $M_\nu$ or $M_{\rm GUT}$. Because of this largely model-dependent scalar sector we will not study gauge coupling unification in detail, but simply assume that there are suitable threshold correction at $M$ and $M_\nu$ that lead to unification around $M_{\rm GUT}$ (it might be necessary to embed our framework into a supersymmetric setup for this purpose). This approach is justified mainly by phenomenology, since our model makes definite predictions for the neutrino sector that can be tested in the near future. 

According to this bottom-up spirit, we first allow only SM singlets $s$ and doublets $h,h^c$ in the ${\bf 27}_H$, ${\bf 351^\prime}_H$ and ${\bf 78}_H$ to take VEVs, where the singlet VEVs are large, i.e ${\cal O} (M_{\rm GUT})$, ${\cal O} (M_\nu)$ or ${\cal O} (M)$, and the SU(2)$_L$ breaking VEVs are at most of the order of the electroweak scale. The complete list of fields with the SM quantum numbers of $s,h,h^c$ contained in the ${\bf 27}_H$, the ${\bf 78}_H$ and ${\bf 351^\prime}_H$ can be found in the Appendix. A second requirement on the scalar VEVs comes from imposing the so-called Barr criteria, which ensure that the low-energy quark mass matrices have real determinants at tree-level. With the shorthand notation for the fermions in Eq.~\eqref{eq:27}, $t = {\bf 10}_{\bf 16}, \overline{f} = {\bf \overline{5}}_{\bf 16}, \overline{F} ={\bf \overline{5}}_{\bf 10}, F = {\bf 5}_{\bf 10}$, the Barr criteria require that 
\begin{itemize}
\item $i$) No SU(2) breaking mass terms for $t - \overline{F}$ are present 
\item  $ii$) Only mass terms for  $\overline{f} - F$ are  complex  
\end{itemize}
If these criteria are fulfilled, one can easily check that the resulting down-quark mass matrix has a real determinant 
, but has entries  that are in general complex 
 and thus can provide the CKM phase. Decomposing the Lagrangian in Eq.~\eqref{L1} under SU(5), one can see from criterion $i$) that the VEVs of the fields $h^c_{\bf 27,16 , \overline{5}}, h^c_{\bf 351, 144, \overline{45}},  h^c_{\bf 351, 144, \overline{5}}$ have to vanish (the subscripts denote the quantum numbers under E$_6$, SO(10) and SU(5), see Appendix for details). Criterion $ii$) implies that only the singlet VEVs $s_{\bf 27, 16, 1} , s_{\bf 351, 144, 24}, $ are complex.
 
\begin{figure}[t]
\begin{center}
\includegraphics[scale=0.3]{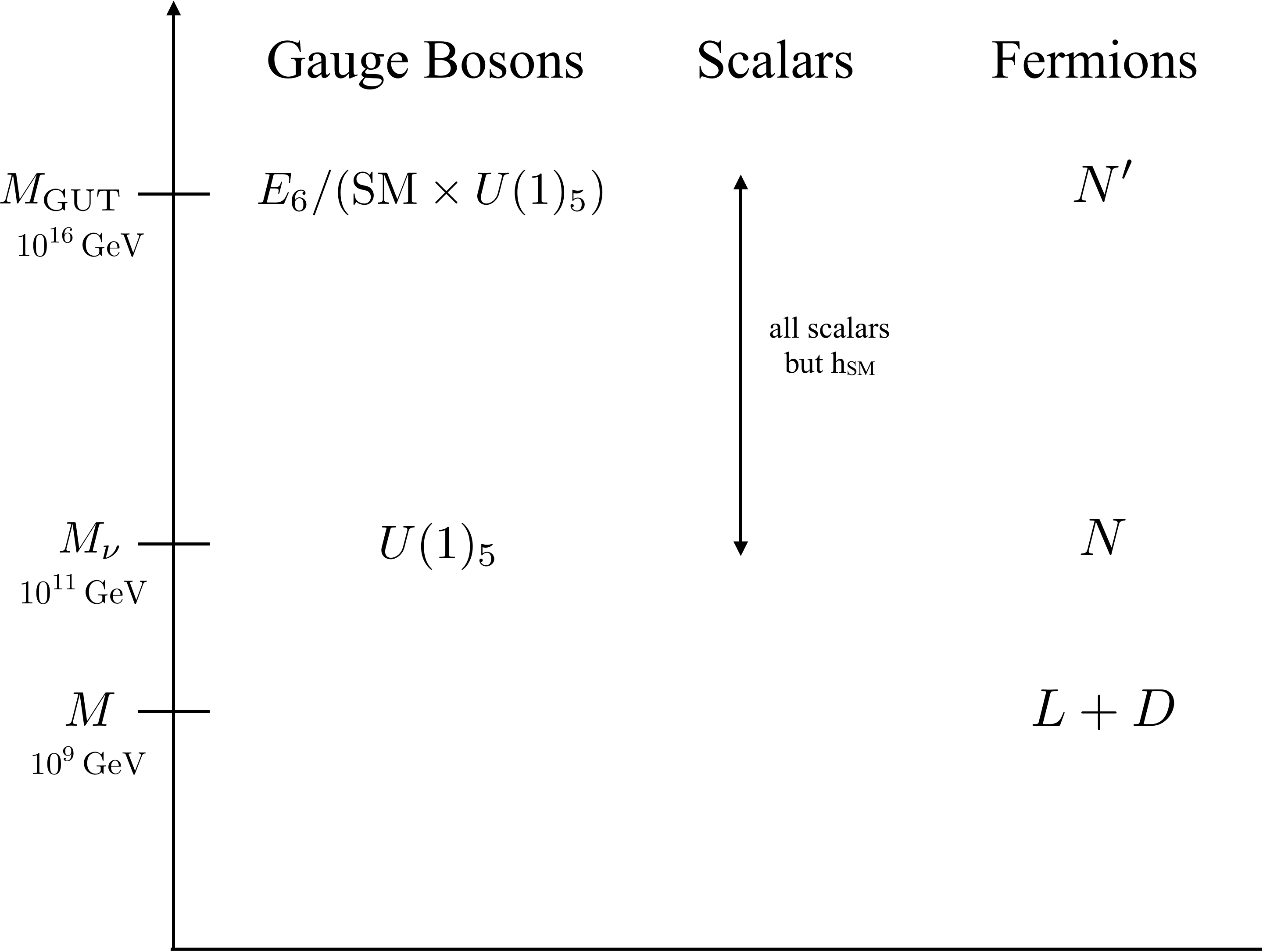}
\caption{\label{scales} Sketch of heavy particle scales, see text for details.}
\end{center}
\end{figure}

 Apart from imposing these conditions on the VEVs, which have to be fulfilled to high degree in order to solve the strong CP problem,  we set some electroweak VEVs to zero that merely lead to sub-leading corrections or can be absorbed into other VEVs. Moreover, for simplicity we also assume that the singlet VEVs giving rise only to neutrino masses are real, although they are not directly constrained by the Barr criteria. As we will discuss below, a complex phase in those VEVs would only affect the Majorana phases, not the Dirac CP phase. We therefore assume the following VEVs in the scalar sector:
\begin{align}
\langle h_{\bf 27,10, 5} \rangle & = v_{u1} \, , & \langle h_{\bf 351, 10,5}  \rangle & = v_{u2} \, , & \langle h^c_{\bf 27, 10, \overline{5}} \rangle & = v_{d1} \, , & \langle h^c_{ \bf 351, 10, \overline{5}} \rangle & = v_{d2} \, ,  \nonumber \\
\langle s_{\bf 27, 16, 1} \rangle & \equiv V^c_{10} \, , &  \langle s_{\bf 351, 144, 24} \rangle & \equiv V^c_{5} \, , &
\langle s_{\bf 27, 1, 1} \rangle & \equiv V_{6}   \, , & \langle s_{\bf 351, 54, 24} \rangle & \equiv V_{5}   \, , \nonumber \\
\langle s_{\bf 351, \overline{126}, 1} \rangle & \equiv V_{10}/2  \, , & \langle s_{\bf 351, 1,1} \rangle & \equiv \tilde{V}_{6}/2 \, , & \langle s_{\bf 351, \overline{16}, 1}  \rangle & \equiv V_{10}^\prime \, , & \langle s_{\bf 78, 45, 24}  \rangle & \equiv \tilde{V}_{5} \, ,  
\label{vevdef}
\end{align}
where all VEVs are real and positive except $V_{10}^c, V_5^c$, of which at least one is complex. Here the subscripts $6,5,10$ denotes the breaking of E$_6$, SU(5) and SO(10), respectively. Apart from the electroweak VEVs, which are all of the order of the weak scale $v = 174 \GeV$, there are three heavy scales $M$, $M_\nu$ and $M_{\rm GUT}$, which set the order of magnitude of the singlet VEVs as
\begin{gather}
 V_6 \sim V_5  \sim  |V_5^c| \sim  |V_{10}^c| \equiv M \sim  10^{9} \GeV \, ,  \nonumber \\ 
V_{10}^\prime \sim V_{10}  \equiv M_\nu \sim 10^{11} \GeV \, , \nonumber \\ 
\tilde{V}_{6} \sim \tilde{V}_5 \equiv M_{\rm GUT} \sim  10^{16} \GeV \, .
\label{singletscales}
\end{gather}
The VEVs determine the breaking pattern of E$_6$ to the SM and set the scale of heavy gauge boson and fermion masses. At $M_{\rm GUT}$ the VEVs  $\tilde{V}_{6}$ and  $\tilde{V}_5$ break $E_6$ to $G_{\rm SM} \times $U(1)$_5$, while the residual U(1)$_5$ factor is broken at $M_\nu$ by $V_{10}$ and $V_{10}^\prime$.  Via the Yukawa couplings in Eq.~\eqref{L1} the VEVs  in Eq.~\eqref{vevdef} generate all fermion masses, which we parametrize as
\begin{align}
\left( m_{10} \right)_{ij}  & =  Y_{27, ij} v_{u1} + Y_{351^\prime, ij}   v_{u2}  \, , & \left( m_5 \right)_{ij}  & =  Y_{27, ij} v_{d1} + Y_{351^\prime, ij}  v_{d2}    \, ,  \nonumber \\  
\left( M_{fF} \right)_{ij} & =Y_{27, ij} V_{10}^c + Y_{351^\prime, ij}  V_{5}^c  \, , &
  \left( M_{FF} \right)_{ij}  & =Y_{27, ij} V_6 + Y_{351^\prime, ij}  V_5    \, , \end{align}
  and
  \begin{align}
  \left( M_{NN } \right)_{ij}  & =  Y_{351^\prime, ij}  V_{10}    \, ,  &     \left( M_{N^\prime N^\prime} \right)_{ij}  & = Y_{351^\prime, ij} \tilde{V}_6 & \left( M_{NN^\prime} \right)_{ij}  & =  Y_{351^\prime, ij}  V_{10}^\prime   \, .
 \label{massdef}
\end{align}
Note that all mass matrices are real and symmetric except $M_{fF}$. Neglecting Clebsch-Gordon coefficients, the fermion masses can be written in SU(5) notation as 
\begin{align} \label{eq:fermionmasses}
{\cal L}_{\rm mass} & = t_i t_j (m_{10})_{ij}  + t_i \overline{f}_{j}  (m_5)_{ij} \nonumber\\
&  + \overline{f}_{i} N_{j} (m_{10})_{ij}    + \overline{F}_{i} N^{\prime}_{j} (m_{10})_{ij} + F_{i} N^{ \prime}_{j} (m_5)_{ij}  \nonumber\\
& + \overline{f}_{i} F_{j} (M_{fF})_{ij}  + \overline{F}_{i} F_{j} (M_{FF})_{ij}  \nonumber\\
&  + \frac{1}{2} N^{ }_{i} N^{ }_{j} (M_{N N})_{ij}  +  \frac{1}{2} N^{ \prime}_{i} N^{ \prime}_{j} (M_{N^\prime N^\prime})_{ij} +  \frac{1}{2} N_{i} N^{ \prime}_{j} (M_{NN^\prime})_{ij}   +{\rm h.c.}, 
\end{align}
where we have introduced the shorthand $N = {\bf 1_{16}}, N^\prime = {\bf 1_{1}}$. 
The first and third line generate masses for quarks and charged leptons, while the second and fourth line are responsible for neutrino masses. The first two lines comprise weak scale SU(2) breaking masses, while the last two lines are heavy mass terms for vector-like fields from the singlet VEVs in Eq.~\eqref{singletscales}. In particular the heavy RH down quarks and LH lepton doublets get a mass at $M$, while the heavy RH neutrinos $N^\prime$ and $N$ get a mass at $M_{\rm GUT}$ and $M_\nu$, respectively. Together with the scalars at $M_\nu$, we have thus fixed the mass scales of all heavy fields (apart from the  additional hierarchies for heavy fermions from hierarchical Yukawa couplings), which we summarize in Fig.~\ref{scales}. 

We will now first neglect the weak scale VEVs and diagonalize the heavy sector given by the last two lines above. In this way we can identify the linear combination of $\overline{f}$ and $\overline{F}$ that remains light and determine the SM quark and charged lepton masses. Similarly we can integrate out the heavy neutrino mass eigenstates to  obtain light neutrino masses. 
\subsection{Quark and Charged Lepton Sector}
We first derive the light mass matrices in SU(5) language and include Clebsch-Gordon coefficients later on. We begin by rewriting $\overline{f}_i,\overline{F}_i$ in terms of light fields $\overline{f}_{Li}$ and heavy fields $\overline{F}_{Hi}$ (that get a mass with $F_i$) with the ansatz 
\begin{align}
\overline{f} & = a_{f} \cdot \overline{f}_L + A_{f} \cdot \overline{F}_H \, , & \overline{F} & = a_{F} \cdot \overline{f}_L + A_{F} \cdot \overline{F}_H \, ,
\end{align}
with some $3 \times 3$ matrices $a_{f}, A_{f}, a_{F}, A_{F}$. Requiring canonically normalized kinetic terms  gives three conditions
\begin{align}
a_{f}^\dagger a_{f} + a^\dagger_{F} a_{F} & = 1_3 \, , & A_{f}^\dagger A_{f} + A^{\dagger }_{F} A_{F}  & = 1_3 \, , & a_{f}^\dagger A_{f} + a_{F}^\dagger A_{F} & = 0 \, , 
\end{align}
and imposing that the light field $ \overline{f}_{L}$ has no mass term with $F$ yields a fourth condition
\begin{align}
a_{f}^T M_{fF} + a_{F}^T M_{FF} = 0 \, .
\end{align}
One can now solve these four equations, but since we are only interested in light fields we need only $a_f$ and $a_F$, which are given by:
\begin{align}
a_f & = \left[ 1_3 + Z^\dagger Z \right]^{-1/2} \, , & a_F & = - Z \cdot a_f \, , & Z & = \left[ M_{fF} (M_{FF})^{-1} \right]^T  \, .
\end{align}
The quark Lagrangian in terms of light fields $ \overline{f}_L$ is 
\begin{align}
{\cal L} & = t_{i} t_{j} (m_{10})_{ij} + t_{i} \overline{f}_{Lj}  (m_5^{\rm eff})_{ij}  + \cdots \end{align}
with 
\begin{align}
m_5^{\rm eff}  & = m_5 \cdot a_f \,  .
\end{align}
Note that while $m_{10}$ and $m_5$ are real symmetric matrices, $a_f$ is hermitian and therefore carries a complex phase into the light Yukawa matrix $m_5^{\rm eff}$ whose determinant is nevertheless real. 

Upon including Clebsch-Gordon coefficients, quark and charged lepton masses $M_i$ are finally given by   
\begin{align}  \label{eq:quarkandleptonmasses}
M_u & = m_u \,, & M_d & = m_d \cdot a_d \,, & M_e & = a_e^T \cdot m_e \,, 
\end{align}
with the real symmetric matrices $m_i$
\begin{align}
m_u & = H r_{\beta 1} + F r_{\beta 2} \, , \\
m_d & = H + F \, , 
\\
m_e & = H - 3  F \, , 
\end{align}
and the hermitian matrices $a_i$
\begin{align}
a_d & = \left[ 1 + Z_d^\dagger Z_d \right]^{-1/2} \, , & Z_d & = r_{10,6} \left( H + F r_{5,6} \right)^{-1} \left( H + F  c_{5,10} \right) \,  ,\\
a_e & = \left[ 1 + Z_e^\dagger Z_e \right]^{-1/2} \, , & Z_e & = r_{10,6} \left( H - \frac{3}{2} F r_{5,6} \right)^{-1} \left( H -   \frac{3 }{2 } F c_{5,10} \right) \, .
\end{align}
The masses depend on the two real symmetric matrices $H$ and $F$ (where we have chosen $F$ to be diagonal)
\begin{align}
H & \equiv Y_{27} v_{d1} \, , \\
F & \equiv Y_{351^\prime}  v_{d2}   \, , 
\end{align}
and five VEV ratios $r_{\beta 1}, r_{\beta 2}, r_{10,6}, r_{5,6}, c_{5,10}$
\begin{gather}
r_{\beta 1}  \equiv \frac{v_{u1}}{v_{d1}} \, ,  \qquad \qquad \qquad r_{\beta 2}  \equiv \frac{v_{u2} }{v_{d2}} \, , \nonumber \\
r_{10,6}  \equiv \frac{|V_{10}^c|}{V_{6}}  \, ,   \qquad \qquad  r_{5,6}  \equiv \frac{V_{5}}{V_{6}} \frac{v_{d1} }{v_{d2} }     \, ,   \qquad \qquad c_{5,10}  \equiv \frac{V_{5}^c}{V_{10}^c}   \frac{v_{d1} }{v_{d2}} \, . 
\label{defratios}
\end{gather}
Since only the ratio $c_{5,10}$ is complex, the quark and charged lepton sectors depend in total on 6+3+5 = 14 real parameters and 1 complex phase. As we will see later, SM fermion masses and mixings can be reproduced if all ratios are ${\cal O}(1)$. \subsection{Neutrino Sector}
The heavy singlet mass terms are given by
\begin{align}
{\cal L}_{N,N^\prime} & =  \frac{1}{2} \left( N_i , N^{ \prime}_i \right) {\cal M}_{N,ij} \begin{pmatrix} N_j \\ N^{ \prime}_j 
\end{pmatrix}  \, , 
\end{align}
with the heavy $6 \times 6$ neutrino mass matrix 
\begin{align}
{\cal M}_{N,ij} = \begin{pmatrix} M_{N N, ij}  & M_{N N^\prime, ij} \\ M_{N N^\prime, ij} & M_{ N^\prime N^\prime, ij} \end{pmatrix} \, .
\end{align}
Note that the $3 \times 3$ sub-matrices defined in Eq.~(\ref{massdef}) are all real and proportional to $Y_{351^\prime}$ that we have chosen to be diagonal. Therefore ${\cal M}_{N,ij}$ is a $2 \times 2$ block matrix of diagonal (and real) $3 \times 3 $ matrices. In order to diagonalize it, we therefore need only to diagonalize the real symmetric $2 \times 2$ matrix ${\cal M}$ defined by
\begin{align}
{\cal M} = \begin{pmatrix}  V_{10} & V_{10}^\prime \\ V_{10}^\prime &  \tilde{V}_{6} \end{pmatrix} \, .
\end{align}
The eigenvalues of this matrix set the scale of light neutrino masses via the Type-I seesaw mechanism. Since we are taking strongly hierarchical VEVs $V_{10} \sim V_{10}^\prime \ll \tilde{V}_6$, we can neglect the off-diagonal entries and the seesaw contribution of $N^{ \prime}$, and integrate out $N$ using the couplings of the light neutrinos $\nu_L$ contained in $\overline{f}_L$ 
 \begin{align}
{\cal L} & =  \nu_L^T  a_e^T  \left( H r_{\beta1} - 3 F r_{\beta2} \right)  N    +{\rm h.c.}  
\end{align}
Light Majorana neutrino masses $m_\nu$ defined as 
$
{\cal L}_{\nu}  = - 1/2  \, m_{\nu, ij} \nu_{L,i} \nu_{L,j}  + {\rm h.c.} \, ,  
$ 
are given by \begin{align} \label{eq:neutrinomassses}
m_\nu & \approx   r_{\eps}  \left[  a_e^T \cdot \left( H r_{\beta1} - 3 F r_{\beta2} \right)  \cdot F^{-1} \cdot \left( H r_{\beta1} - 3 F r_{\beta2} \right) \cdot  a_e \right] \, ,
\end{align}
where have introduced the VEV ratio
\begin{align}
r_{\eps} & \equiv \frac{v_{d2}  }{V_{10}} \lll 1 \, .
\end{align}
Here we are neglecting contributions from the other heavy neutrinos at $M_{\rm GUT}$ and the heavy mixing due to the presence of $V_{10}^\prime$, but we have checked that including these correction only amounts to tiny corrections to Eq.~\eqref{eq:neutrinomassses}, of the order of the naive scale suppression factor  $M_\nu/M_{\rm GUT} \sim 10^{-5}$. Therefore possible phases in $V_{10}^\prime$ and $\tilde{V}_6$ would not affect the light neutrino sector, while a possible phase in $V_{10}$ would only lead to an overall phase of the light neutrino mass matrix, and thus only affects Majorana phases, not the Dirac CP phase.

Counting parameters, we see that the neutrino sector depends on a single additional real VEV ratio  $r_{\eps}$ compared to the quark and charged lepton sector. Therefore we have in total 15 relevant real parameters + 1 phase to describe the measured 17 + 1 SM parameters: 9 quark and charged lepton masses, 2 neutrino mass differences, 6 mixing angles and 1 CKM phase.  This means that there are two predictions that make the fit of the parameters to experimental data non-trivial. Moreover the model makes definite predictions for yet unmeasured observables in the neutrino sector (Dirac phase, two Majorana phases, overall neutrino mass scale, effective scale for neutrinoless double beta decay) and is therefore testable. We discuss the fit and these predictions in the next section.

\section{Fit to Fermion Masses and Mixings}

\begin{small}
\begin{table} [t]
\centering
\begin{math}
\begin{array}{|c|c||c|c|}
\hline
 \multicolumn{4}{|c|}{\text{Fermion observables at the electroweak scale $\mu=M_Z$} }\\
\hline
 m_d(\text{MeV}) &2.75 \pm 0.29 	&\Delta_{12}(\text{eV$^2$})&(7.50 \pm 0.18)\times10^{-5} \\ 
 m_s(\text{MeV}) &54.3 \pm 2.9		&\Delta_{31}(\text{eV$^2$})&(2.52 \pm 0.04)\times
10^{-3}\\	
 m_b(\text{GeV}) & 2.85 \pm 0.03				& \sin  \theta _{12}^q & 0.2254 \pm 0.0007 \\
 m_u(\text{MeV}) & 1.3 \pm 0.4				& \sin  \theta _{23}^q & 0.0421 \pm 0.0006 \\
 m_c(\text{GeV}) & 0.627 \pm 0.019 		& \sin  \theta _{13}^q & 0.0036 \pm 0.0001 \\
 m_t(\text{GeV}) &171.7\pm 1.5				& \sin ^2 \theta _{12}^l & 0.306\pm 0.012 \\
 m_e(\text{MeV}) &0.4866\pm 0.0005 		&\sin ^2 \theta _{23}^l&0.441 \pm 0.024  \\
 m_{\mu }(\text{MeV})&102.7\pm 0.1 		& \sin ^2\theta _{13}^l & 0.0217 \pm 0.0008 \\
 m_{\tau }(\text{GeV})&1.746\pm 0.002	&\delta_{\rm CKM}&1.21 \pm 0.05\\
\hline
\end{array}
\end{math}
\vspace{0.5cm}
\caption{SM input parameters at the electroweak scale, where quark and lepton masses and the quark mixing parameters are taken from Ref.~\cite{Antusch:2013jca}, and neutrino mixing parameters  from Ref.~\cite{Esteban:2016qun} for Normal Ordering (NO). As explained in the text, we use a $0.1 \%$ uncertainty for the charged lepton masses in the fitting procedure. To simplify the fitting procedure, we used for all observables the arithmetic average of the errors when not symmetric. 
}
\label{tab:smmasses}
\end{table}
\end{small}

In order to verify whether the 17+1 fermion observables of the Standard Model (see Table~\ref{tab:smmasses} for our input) can be successfully reproduced in our model, we have performed a fit of the matrices $H, F$ and the six VEV ratios $r_{\beta 1} ,  r_{\beta 2},r_{10,6} , r_{5,6}, c_{5,10} , r_{\epsilon} $ using Eq.~(\ref{eq:quarkandleptonmasses}) and Eq.~(\ref{eq:neutrinomassses}), corresponding to 15 real parameters + 1 complex phase. The fit was done using the Metropolis-Hastings algorithm \cite{Metropolis53,Hastings70} following a top-down approach. The parameters were chosen randomly at $Q = 10^{16} \GeV$ and used as boundary conditions for the Yukawa RGEs, which were then solved numerically using REAP \cite{Antusch:2005gp}. Afterwards, the computed values of the observables at the electroweak scale $M_Z$ were compared with the experimental values (following earlier studies, we assume a $0.1 \%$ uncertainty for the charged lepton masses \cite{Dueck:2013gca,Meloni:2014rga}, because otherwise a numerical fit would be very challenging). The quality of a fit point is determined by
\begin{equation} \label{eq:chisquare}
    \chi^2 \equiv \sum_{i=1}^n \left(\frac{{\cal O}^{\rm exp}_i-{\cal O}^{\rm fit}_i}{\sigma^{\rm exp}_i}\right)^2 \, ,
\end{equation}
where ${\cal O}^{\rm exp}_i$ denotes the experimental value of the observable ${\cal O}_i$, $\sigma^{\rm exp}_i$ its experimental error and ${\cal O}^{\rm fit}_i$ the corresponding fit value. 

\vspace{0.3cm}

\noindent Despite the overdetermination\footnote{There must be two relations involving just SM observables, however due to the highly non-trivial dependence on the fundamental parameters we were not able to find analytical expressions for these relations.} we find a perfect fit with $\chi^2/{\rm dof} \approx 0.9$ for NO in the neutrino sector, with corresponding model parameters at the GUT scale given by
\begin{gather}
    H=\left(
\begin{array}{ccc}
 -0.00814 & 0.0292  & -0.0894  \\
0.0292 & -0.217 & 2.49 \\
 -0.0894  & 2.49 &  -12.8 \\
\end{array}
\right) \, , \qquad
F=\left(
\begin{array}{ccc}
 -0.00248  & 0. & 0. \\
 0. & 0.0489  & 0. \\
 0. & 0. & 30.7 \\
\end{array}
\right) \notag \\ \nonumber \\
r_{\beta 1} = -1.28 \, ,  \quad r_{\beta 2}  = 2.26 \, , \quad  r_{10,6}  = 2.21 \, , \quad  r_{5,6}  = -0.433     \, ,  \nonumber \\
 c_{5,10}  = 2.20 \cdot e^{1.60 \, i } \, , \quad     r_{\epsilon}=  1.73  \cdot 10^{-10} \, ,
\end{gather}
and the fitted standard model fermion observables are summarized in Table~\ref{tab:fitresults}.
\begin{small}
\begin{table} [ht]
\centering
\begin{math}
\begin{array}{|c|c|c||c|c|c|}
\hline
 \multicolumn{6}{|c|}{\text{Fit result at the electroweak scale $\mu=M_Z$} }\\
\hline 
 &  \text{fit} &  \text{pull} &   &  \text{fit} &  \text{pull} \\ \hline 
 m_d(\text{MeV}) &3.44 & -2.4 	&\Delta_{12}(\text{eV$^2$})&7.39 \times 10^{-5} & 0.63 \\ 
 m_s(\text{MeV}) &50.4 & 1.4		&\Delta_{13}(\text{eV$^2$})&-0.76 \times
10^{-3} & -0.19\\	
 m_b(\text{GeV}) & 2.85 & 0.27				& \sin  \theta _{12}^q & 0.225 & 0.56 \\
 m_u(\text{MeV}) & 1.32 & -0.08				& \sin  \theta _{23}^q & 0.0414 & 0.1 \\
 m_c(\text{GeV}) & 0.63 & -0.07 		& \sin  \theta _{13}^q & 0.0035& 1.1 \\
 m_t(\text{GeV}) &171.58 & 0.08				& \sin ^2 \theta _{12}^l & 0.302 & 0.37 \\
 m_e(\text{MeV}) &0.486& 0.15 		&\sin ^2 \theta _{23}^l&0.405 & 1.5  \\
 m_{\mu }(\text{MeV})&102.76 & -0.61 		& \sin ^2\theta _{13}^l & 0.022 & -0.26 \\
 m_{\tau }(\text{GeV})&1.746 & -0.04	&\delta_{\rm CKM}&1.13 & 1.5\\
\hline
\end{array}
\end{math}
\caption{Result of the fitting procedure, as described in the text. The pull of a fit value ${\cal O}^{\rm fit}_i$ is defined as $\text{pull}({\cal O}^{\rm fit}_i)=\left( {\cal O}^{\rm exp}_i-{\cal O}^{\rm fit}_i \right)/\sigma^{\rm exp}_i$, where $\sigma^{\rm exp}_i$ is the corresponding experimental error and ${\cal O}^{\rm exp}_i$ the experimental value as given in Table~\ref{tab:smmasses}. } \label{tab:fitresults}
\end{table}
\end{small}
Using the above fit parameters, we can also make predictions for the neutrino Dirac phase, Majorana phases and neutrino mass observables. Experiments that are sensitive to the absolute neutrino mass scale, like the KATRIN \cite{Osipowicz:2001sq},  MARE \cite{Monfardini:2005dk} , Project 8 \cite{Monreal:2009za}, 
or ECHo \cite{Blaum:2013pfu} experiments, measure the effective mass $m_\beta$ defined by
\be \label{eq:mbeta}
m_\beta = \sqrt{\sum |U_{ei}|^2 \, m_i^2} \, . 
\ee\\
In contrast, cosmology probes the sum of neutrino masses $\Sigma = \sum m_i$ and neutrinoless double beta decay experiments, like, for example, the GERDA \cite{Agostini:2013mzu}, EXO-200 \cite{Albert:2014awa} or KamLAND-Zen \cite{Gando:2012zm} experiments, tests the ``effective Majorana mass"
\be \label{eq:meff}
m_{\beta \beta} = \left| \sum U_{ei}^2 \, m_i \right|  \, .
\ee
To give an estimate for the robustness of these predictions in our model, we  numerically considered perturbations around the best fit point that reproduce the standard model fermion observables with $\chi^2/{\rm dof} \lesssim 10$. Our predictions and the resulting ranges for the above mass observables and Dirac and Majorana phases are summarized in Table~\ref{tab:predictionsandboundsneutrinomasses}. While the mass observables are all far below current and future sensitivity, we obtain a quite narrow range for the Dirac phase $\delta \in \left[ 154, 157 \right]^\circ$, which might be verified or excluded with upcoming data coming from neutrino oscillations, for example at Hyper-Kamiokande~\cite{HK2} or DUNE~\cite{Dune}. 
\begin{table}[]
\centering
\begin{tabular}{|c|c|c|c|c|c|c|}
\hline
                       & $m_\beta$ {[}meV{]} & $\Sigma$ {[}meV{]} & $m_{\beta \beta}$ {[}meV{]} &$ \delta$ {[}$ ^\circ ${]}& $\varphi_1$ {[}$^\circ${]} & $\varphi_2$ {[}$^\circ${]} \\ \hline & & & & & & \\[-1em]
Prediction &  $8.8 \pm 0.5   $  &   $ 59 \pm 3  $        &   $ 1.8 \pm 0.1 $       & $157 \pm 3 $  & $187 \pm 4$  &  $159 \pm 5 $      \\ \hline
Current bound  & $\lesssim 2000 $~\cite{Patrignani:2016xqp}      &$\lesssim 230$~\cite{Patrignani:2016xqp, Ade:2015xua}    & $200$~\cite{Capozzi:2017ipn, KamLAND-Zen:2016pfg}     & - &  - &  - \\ \hline
\end{tabular}
\caption{Predicted values and current bounds for the neutrino observables. The current bounds were taken from Ref. \cite{Pas:2015eia}. As explained in the text, the ranges shown here correspond to perturbations of the best fit point with $\chi^2/{\rm dof} \lesssim 10$.}
\label{tab:predictionsandboundsneutrinomasses}
\end{table}

Finally we comment on the remaining free parameters that are left undetermined by the fit to masses and and mixings. From 12 VEVs (see Eq.~\eqref{vevdef}) six are determined by the fit and one by the electroweak scale ($v^2 = v_{u1}^2 + v_{d1}^2 + v_{u2}^2 + v_{d2}^2$). From the remaining 5 VEVs, $V_{10}^\prime$ does not affect the neutrino sector given the hierarchy $V_{10}^\prime \sim V_{10} \ll \tilde{V}_6 $, so for simplicity we set $V_{10}^\prime = V_{10} = M_\nu = v_{d2}/r_\eps \sim 10^{11} \GeV$ without any impact on the spectrum.  We are then left with four VEVs that are free parameters, which we take as $v_{d1}, |V^c_{10}|, \tilde{V}_{6}, \tilde{V}_{5} $. As the two latter VEVs control the mass of heavy gauge bosons, they are bounded from below by proton decay constraints, and we take $\tilde{V}_{6} = \tilde{V}_{5} \equiv M_{\rm GUT} \sim 5 \cdot 10^{16} \GeV$. The VEV
$v_{d1}$ is mainly bounded by requiring perturbative Yukawa couplings and does not have a big impact on the spectrum, and for the sake of explicitness we fix  $v_{d1} = 70 \GeV$. The remaining scale $M$ is bounded from above by neutron EDM constraints, which require that the loop corrections to the effective $\overline{\theta}$ parameter remain sufficiently small. As we will discuss in the next section, these higher-loop corrections are sufficiently suppressed if $M \sim 10^9 \GeV$.

\section{Loop Contributions to $\overline{\theta}$}

In Nelson-Barr models $\overline{\theta}$ vanishes at tree-level by construction, but is generated at loop-level due to higher order corrections to the effective Yukawa couplings. Therefore care has to be taken to ensure that such corrections are sufficiently small in order to have $\overline{\theta} < 10^{-10}$. The form of these (finite) corrections has already been discussed to large extent in the literature for the original Nelson model~\cite{Nelson2} and in more general setups~\cite{Vecchi}. It turns out that such corrections are in general suppressed by loop factors and small Yukawa couplings and/or small mass ratios. While the contributions suppressed by Yukawas are always negligibly small in Nelson-Barr type models where only the RH down quarks mix with heavy fields~\cite{Vecchi}, the contributions sensitive to UV physics are suppressed by ratios of the heavy RH down quark masses over heavy gauge boson or heavy scalar masses~\cite{Nelson2, Branco1}. Thus they can be made sufficiently small by lowering the mass scale of RH down quarks $M$, which in our setup is a free parameter. In this section we (conservatively) estimate the leading corrections involving heavy gauge bosons and scalars using a spurion analysis, showing that $M \sim 10^{9}$ is enough to render $\overline{\theta} \lesssim 10^{-10}$. 

In this spirit we work with a simplified Lagrangian before going  to the light-heavy mass basis, cf. Eq.~\eqref{eq:fermionmasses}
\begin{align}
{\cal L} = q^T \lambda_u u \, h + q^T \lambda_d d \, \tilde{h} + d^T m_{dD} \overline{D} + d^T \kappa_{10} \overline{D} S_{10} + d^T \kappa_{5} \overline{D} S_{5} + D^T M_D \overline{D} + {\rm h.c.}
\end{align}
where we also included scalar couplings $\kappa_{10}, \kappa_5$ and all masses and couplings are real symmetric $3 \times 3$ matrices except $m_{dD}$ that is complex symmetric 
\begin{align}
\lambda_u & = \frac{1}{v} \left( H r_{\beta 1} + F r_{\beta 2} \right) \, , &
\lambda_d & = \frac{1}{v} \left( H  + F   \right) \, , \nonumber \\
m_{dD} & = \frac{V_{10}^c}{v_{d1}} \left( H + F c_{5,10} \right) \, , &
M_{D} & = \frac{|V_{10}^c|}{v_{d1} r_{10,6}} \left( H + F r_{5,6} \right) \, , \nonumber \\
\kappa_{10} & = \frac{H}{v_{d1} \sqrt2} \, , & \kappa_{5} & = \frac{F}{v_{d2} \sqrt2} \, .
\end{align}
 We are now interested in loop corrections to the Yukawa couplings $\lambda_{u,d}$ that we write as 
\begin{align}
\lambda_u^{\rm tot} & = \lambda_u + \Delta \lambda_{u} \, , & \lambda_d^{\rm tot} & = \lambda_d + \Delta \lambda_{d} \, .
\end{align}
The effective SM Yukawa couplings $y_{u,d}$ are given by (cf. Eq.~\eqref{eq:quarkandleptonmasses})
\begin{align}
y_u & = \lambda_u + \Delta \lambda_{u} = \lambda_u \left[1 + \lambda_u^{-1} \Delta \lambda_{u}  \right]\, , \nonumber \\
 y_d  & = \left( \lambda_d   +  \Delta \lambda_{d} \right) a_d =  \lambda_d a_d \left[ 1 + a_d^{-1} \lambda_d^{-1} \Delta \lambda_{d} a_d \right] \, .
\end{align}
Therefore the effective $\theta$ parameter is 
\begin{align}
\overline{\theta} & = \arg \det y_u  y_d  =   { \rm Im} \,  {\rm tr} \log y_u  +  { \rm Im} \,  {\rm tr} \log y_d  \, , \nonumber \\
& =  { \rm Im} \,  {\rm tr} \log  \left[1 + \lambda_u^{-1} \Delta \lambda_{u}  \right] + { \rm Im} \,  {\rm tr} \log  \left[1 + a_{d}^{-1} \lambda_d^{-1} \Delta \lambda_{d} a_{d} \right] \, , 
\end{align} 
where have used  that $\lambda_{u}, \lambda_{d}$ are real and $a_d$ is hermitian. Expanding the logarithms we finally get
\begin{align}
\overline{\theta} & =   \sum_n  \frac{1}{n}  \left[ { \rm Im} \,  {\rm tr}\left(  \lambda_u^{-1} \Delta \lambda_{u}  \right)^n +  { \rm Im} \,  {\rm tr}\left(  \lambda_d^{-1} \Delta \lambda_{d}  \right)^n  \right] =  { \rm Im} \,  {\rm tr}\left(  \lambda_u^{-1} \Delta \lambda_{u}  + \lambda_d^{-1} \Delta \lambda_{d}  \right) + \hdots 
\end{align}
We can now estimate the corrections $\Delta  \lambda_{u,d}$ using a spurion analysis. Under unitary field redefinitions $f \to V_f f$ with $f= q,u,d,D,\overline{D}$,  the Lagrangian parameters transform as
\begin{align}
\lambda_u & \to V_q^T \lambda_u V_u \, , & \lambda_d & \to V_q^T \lambda_d V_d \, , \nonumber \\
m_{dD} & \to V_d^T m_{dD} V_{\overline{D}} \, , & M_{D} & \to V_D^T M_{D} V_{\overline{D}}  \, , \nonumber \\
\kappa_{10} & \to V_d^T \kappa_{10} V_{\overline{D}} \, , & \kappa_{5} & \to V_d^T \kappa_{5} V_{\overline{D}} \, , &
\end{align}
Since we can treat the masses $m_{dD}$ as insertions, the covariant expressions involving $m_{dD}$ and $M_D$ that will enter the contributions to $\Delta_{\lambda_u,d}$ can only be of the form  
\begin{gather}
\xi_{\rm IR}  \equiv  (m_{dD})^* (M_D^{-1})^* (M_D^{-1})^T (m_{dD})^T = Z_d^\dagger Z_d\, , \nonumber \\ 
\xi_{\rm V}   \equiv m_{dD}^* m_{dD}^T/M_V^2 \, , \qquad \qquad \xi_{\rm S}   \equiv \kappa_i^*  m_{dD}^T \kappa_i^* m_{dD}^T/M_S^2 \, , 
\end{gather}
where $M_V, M_S$ are the heavy gauge boson and scalar masses. These expressions transform as
\begin{align}
\xi_{\rm IR,V,S} & \to V_q^\dagger \xi_{\rm IR,V,S} V_q \, ,
\end{align}
and are the only quantities that carry complex phases. The heaviest particle in the loop diagram then determines the form of $\xi$, i.e. diagrams with only Higgs scalars, the ``IR" contributions, will involve $\xi_{\rm IR}$, diagrams with heavy vectors $\xi_{\rm V}$ and diagrams with heavy scalars $\xi_{\rm S}$. 

\subsection{IR Contributions}

For IR diagrams $\xi_{\rm IR} \sim {\cal O}(1)$, so one has to sum up all insertions leading effectively to a field redefinition $d \to a_d d$, and thus $\lambda_d \to \lambda_d a_d = y_d$ which is the SM down Yukawa coupling. The spurion analysis is then quite involved, since one has to take into account that the light quark propagators can involve hermitian functions of $y_u y_u^\dagger (y_d y_d^\dagger)^{-1}$ arising from the integration over loop momenta. In Ref.~\cite{Vecchi} it was shown that the leading contribution arises at three loop and is given by
\begin{align}
\Delta_{\rm IR} \overline{\theta} \sim \frac{1}{(4 \pi^2)^3} { \rm Im} \,  {\rm tr} \left[ f_1 y_d^\dagger y_u y_u^\dagger y_d f_2 y_d^\dagger y_d \right] \sim 10^{-16} \, ,
\end{align}  
where the  functions $f_1 \ne f_2$ were taken to as ${\cal O}(1)$ hermitian matrices. Note that the strong suppression is due to the fact that there is no mixing with heavy fields in the up and the LH down sector.

\subsection{Gauge Contributions}
\begin{figure}[h]
\begin{center}
\includegraphics[scale=0.9]{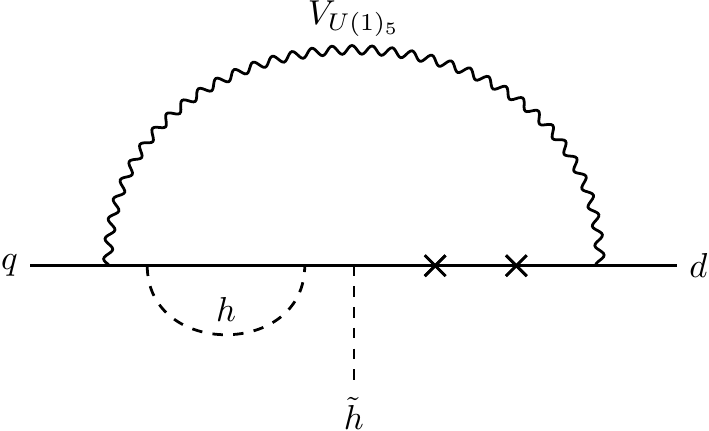}
\caption{\label{2loopgauge} Example of 2-loop diagram with heavy gauge fields.}
\end{center}
\end{figure}

The most dangerous contributions involve the U(1)$_5$ gauge boson, since all heavy gauge fields live at $M_{\rm GUT}$. The spurion analysis is greatly simplified, because now the heavy gauge boson dominates the loop momentum integration, and thus strongly suppresses the contributions from propagators with non-trivial flavor structure. One can show that all one-loop contributions vanish, and that the leading contribution arises from two-diagrams like Fig.~\ref{2loopgauge} that involve additional Higgs loops. 
An estimate gives
\begin{align}
\Delta_{\rm V} \overline{\theta} \sim \frac{ \log{M_f^2/M_V^2}}{(4 \pi^2)^2} { \rm Im} \,  {\rm tr} \left[ \lambda_d^{-1}g_{qq}  \lambda_u \lambda_u^\dagger \lambda_d \xi_{\rm V} g_{dd} \right]  \, ,
\end{align}  
where $g_{qq} = g$ and $g_{dd} = -3 g$ are the couplings of the U(1)$_5$ gauge boson to quarks, $M_V$ is its mass and $M_f$ denotes the heaviest   of the heavy vectorlike down quarks. All numerical values depend only on $M$ (fixing $v_{d1} = 70 \GeV$). Taking $M = 10^{9} \GeV$ implies $M_V = 2 \cdot 10^{12} \GeV$ and $M_f = 10^9 \GeV$, and taking for simplicity the GUT value of the U(1)$_5$ gauge coupling $g \sim  \sqrt{4 \pi \alpha_{\rm GUT}} \sim \sqrt{4 \pi/24}  \approx 0.7$, one finds
\begin{align}
\Delta_{\rm V} \overline{\theta} \sim  3 \cdot 10^{-11} \, .
\end{align}
\subsection{Scalar Contributions}
\begin{figure}[h]
\begin{center}
\includegraphics[scale=0.9]{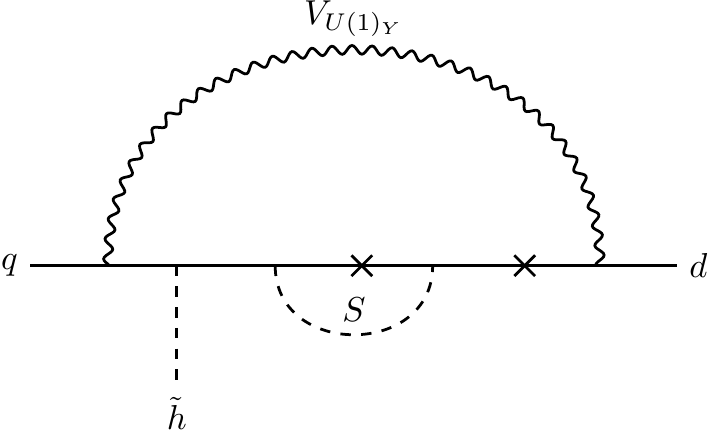}
\caption{\label{2loopscalar} Example of 2-loop diagram with heavy scalar fields.}
\end{center}
\end{figure}

Analogously,  all one loop diagrams with heavy scalars in the loop vanish, and the leading contribution comes from a 2-loop diagram involving a U(1)$_Y$ and scalar loop, shown in Fig.~\ref{2loopscalar}. 
This gives 
\begin{align}
\Delta_{\rm S} \overline{\theta} \sim \frac{ \log{M_f^2/M_S^2}}{(4 \pi^2)^2} { \rm Im} \,  {\rm tr} \left[ \lambda_d^{-1} g_{qq}^Y \lambda_d \xi_{\rm S} g_{dd}^Y \right]  \, ,
\end{align}  
where $g_{qq}^Y = g_Y/6$ and $g_{dd}^Y = g_Y/3$. Note that this term is simply proportional to the imaginary part of the trace of $\xi_S$, which is in general not a hermitian matrix, and arises because we assume an ${\cal  O}(1)$ splitting between the masses of the scalar and pseudoscalar components of the complex fields $S_5$ and $S_{10}$. For degenerate $S_5$ and $S_{10}$, the $S_5$ contribution dominates since it has larger couplings to the heavy fermions, and for $M= 10^9 \GeV$, $M_S = M_\nu = 3 \cdot 10^{11} \GeV$ and $g_Y (10^9 \GeV) \sim 0.4$ one gets 
\begin{align}
\Delta_{\rm S} \overline{\theta} \sim  4 \cdot 10^{-11} \, .
\end{align}

\section{Summary and Conclusions}

To summarize, we have shown that the Nelson-Barr mechanism can be naturally realized in the context of an $E_6$ GUT. The SM fermions are embedded in three generations of E$_6$ fundamentals together with the Nelson-Barr fields (vectorlike RH down quarks), a vectorlike pair of LH lepton doublets  and two RH neutrinos. All heavy mass terms arise from VEVs of scalar fields in E$_6$ representations, whose couplings to fermions are given by two $3 \times 3$ symmetric matrices. CP is imposed on the Lagrangian, and therefore all Yukawa couplings are real. CP is broken spontaneously by two scalar VEVs that mix the chiral SM fermions with the Nelson-Barr fields, through which CP violation is mediated to the low-energy theory. By integrating out the heavy down-quarks, lepton doublets and neutrinos, we have derived analytic formulas for all SM fermion mass matrices, including Majorana neutrino masses. The resulting SM quark mass matrices are real (up sector) and the product of a real and a hermitian matrix (down sector), thus implying that  $\overline{\theta}$ vanishes at tree-level. 

Besides solving the strong CP problem with the Nelson-Barr mechanism, the main benefit of the GUT setup is the predictivity in the fermion sector. The fundamental Yukawa matrices contain just 9 parameters, which together with 6 real VEV ratios and a single complex phase determine the complete SM fermion sector including neutrinos (17 real observables + 1 CKM phase). Surprisingly, we nevertheless obtain a perfect fit with $\chi^2/{\rm dof} \approx 0.9$, implying that there are two relations among SM observables that hold to good precision, but unfortunately we were not able to derive them in closed form due to the complexity of the analytical expressions. Since all low-energy parameters of the model are fixed, we obtain definite predictions for the neutrino sector that makes this model testable in the near future, as shown in Table~\ref{tab:predictionsandboundsneutrinomasses}. Particularly interesting is the prediction of the Dirac CP phase $\delta_{\rm CP} = 157 \pm 3  ^\circ $, which is directly correlated with the CKM phase, and will be verified or excluded by future experiments like  Hyper-Kamiokande~\cite{HK2} or DUNE~\cite{Dune}.

The fit to the fermion sector determines the absolute mass scales only in the neutrino sector, fixing the mass of RH neutrinos at about $M_\nu \sim 10^{11} \GeV$ (there is another SM singlet around the GUT scale $M_{\rm GUT} \sim 5 \cdot 10^{16} \GeV$). The overall mass scale in the heavy RH down and LH lepton sector $M$ is left undetermined, but bounded from above to keep loop corrections to $\overline{\theta}$ sufficiently small. The leading loop contributions from diagrams involving heavy gauge bosons and heavy scalars are suppressed by mass ratios $M^2/M_{V,S}^2$, and we have (conservatively) estimated that $M = 10^9 \GeV$ is enough to render $\overline{\theta} < 10^{-10}$. This essentially fixes all heavy mass scales in our model, as sketched in Fig.~\ref{scales}.

\section*{Acknowledgements}
We thank L.~di Luzio, U.~Nierste and L.~Vecchi for helpful discussions, and L.~di Luzio for useful comments on the manuscript. J.S. and P.T. acknowledge support from the DFG-funded doctoral school KSETA. The Feynman diagrams were drawn using Ti$k$Z-Feynman~\cite{Ellis:2016jkw}.

\appendix
\section{$E_6$ Decomposition}   

\begin{table}[H]
\scriptsize
\centering
\begin{tabular}{|c|c|c|c|}
\hline
$E_6$ & $SO(10) \times U(1)_{10} $ & $SU(5) \times U(1)_{5} \times U(1)_{10}$ & SM \\
\hline
${\bf 27}_F$ & ${\bf 16}_1$ & ${\bf 10}_{1,1} = t$ & $q, u, e $\\
& & ${\bf \overline{5}}_{-3,1} = \overline{f} $ & $d, l$ \\
& & ${\bf 1}_{5,1}$ & $N$\\
&&& \\
& ${\bf 10}_{-2}$ & ${\bf \overline{5}}_{2,-2} = \overline{F}$ & $D, L$ \\
& & ${\bf 5}_{-2,-2} = F$ & $\overline{D}, \overline{L}$  \\
&&& \\ 
& ${\bf 1}_4$ & ${\bf 1}_{0,4}$ &$N^{ \prime}$ \\
&&& \\ 
\addlinespace[-2.4ex]
\hline
${\bf 27}_S$ & ${\bf 16}_1$ & ${\bf 10}_{1,1} $ & \\
& & ${\bf \overline{5}}_{-3,1}  $ & $h^c_{\bf 27, 16, \overline{5}}$ \\
& & ${\bf 1}_{5,1}$ & $s_{\bf 27, 16,1}$ \\
&&& \\
& ${\bf 10}_{-2}$ & ${\bf \overline{5}}_{2,-2} $ &   $h^c_{\bf 27, 10, \overline{5}}$\\
& & ${\bf 5}_{-2,-2} $ &   $h_{\bf 27, 10,5}$ \\
&&& \\ 
& ${\bf 1}_4$ & ${\bf 1}_{0,4}$ &   $s_{\bf 27, 1,1}$ \\
&&& \\
\addlinespace[-2.4ex]
\hline
${\bf 78}_S$ & ${\bf 45}_0$ & ${\bf 24}_{0,0} $ & $s_{\bf 78, 45,24}$  \\
& & ${\bf \overline{10}}_{4,0}  $ &  \\
& & ${\bf 10}_{-4,0}$ & \\
& & ${\bf 1}_{0,0}$ & $s_{\bf 78, 45,1}$ \\
&&& \\
& ${\bf 16}_{-3}$ & ${\bf 10}_{1,-3} $ &  \\
& & ${\bf \overline{5}}_{-3,-3} $ &   $h^c_{\bf 78, 16, \overline{5}}$ \\
& & ${\bf 1}_{5,-3} $ &   $s_{\bf 78, 16,1}$ \\
&&& \\ 
& ${\bf \overline{16}}_{3}$ & ${\bf \overline{10}}_{-1,3} $ &  \\
& & ${\bf 5}_{3,3} $ &   $h_{\bf 78, \overline{16}, 5 }$ \\
& & ${\bf 1}_{-5,3} $ &  $s_{\bf 78, \overline{16},1}$  \\
&&& \\ 
& ${\bf 1}_0$ & ${\bf 1}_{0,0}$ &   $s_{\bf 78, 1,1}$ \\
&&& \\
\addlinespace[-2.4ex]
\hline
${\bf 351}^\prime_S$ & ${\bf 144}_1$ & ${\bf \overline{45}}_{-3,1} $ & $h^c_{\bf 351, 144, \overline{45}}$ \\
& & ${\bf 40}_{1,1}  $ &  \\
& & ${\bf 24}_{5,1}$ & $s_{\bf 351,144,24}$ \\
& & ${\bf 15}_{1,1}$ & \\
& & ${\bf 10}_{1,1}$ & \\
& & ${\bf 5}_{-7,1}$ & $h_{\bf 351, 144, 5}$ \\
& & ${\bf \overline{5}}_{-3,1}$ & $h^c_{\bf 351, 144, \overline{5}}$ \\

&&& \\
& ${\bf \overline{126}}_{-2}$ & ${\bf50}_{-2,-2} $ &  \\
& & ${\bf \overline{45}}_{2,-2} $ &   $h^c_{\bf 351, \overline{126}, \overline{45}}$ \\
& & ${\bf 15}_{6,-2} $ &   \\
& & ${\bf \overline{10}}_{-6,-2} $ &   \\
& & ${\bf 5}_{-2,-2} $ &   $h_{\bf 351, \overline{126},5}$  \\
& & ${\bf 1}_{-10,-2} $ &   $s_{\bf 351, \overline{126}, 1}$  \\
&&& \\ 
& ${\bf 54}_4$ & ${\bf 24}_{0,4}$ &   $s_{\bf 351, 54, 24}$ \\
& & ${\bf \overline{15}}_{4,4}$ &   \\
& & ${\bf 15}_{-4,4}$ &    \\
&&& \\
& ${\bf \overline{16}}_{-5}$ & ${\bf \overline{10}}_{-1,-5}$ &   \\
&  & ${\bf5}_{3,-5}$ &    $h_{\bf 351, \overline{16},5}$ \\
&  & ${\bf 1}_{-5,-5}$ &   $s_{\bf 351, \overline{16},1}$ \\
&&& \\
& ${\bf 10}_{-2}$ & ${\bf \overline{5}}_{2,-2}$ &   $h^c_{\bf 351, 10, \overline{5}}$  \\
& & ${\bf 5}_{-2,-2}$ &   $h_{\bf 351, 10, 5}$  \\
&&& \\
& ${\bf 1}_{-8}$ & ${\bf 1}_{0,-8}$ &   $s_{\bf 351, 1,1}$ \\
&&& \\
\hline
\end{tabular}
\end{table}

\bibliographystyle{JHEP} 
\bibliography{bib}

\end{document}